\documentstyle[epsf]{mn}
\title[The low mass white dwarf companion to Beta Crateris]
{The low mass white dwarf companion to Beta Crateris\thanks{Based on 
observations with the {\it Far Ultraviolet Spectroscopic Explorer}, 
and the {\it Extreme Ultraviolet Spectroscopic Explorer}}}
\author[M.\,R. Burleigh et~al.]
{M.\,R. Burleigh, M.\,A. Barstow, K.\,J. Schenker, A.\,I. Sills, G.\,A. Wynn
\\
{\LARGE  P.\,D. Dobbie and S.\,A. Good
}\\
Department of Physics and Astronomy, University 
of Leicester, University Rd., Leicester, LE1 7RH \\
}

\date{June 25th 2001}

\def\la{\mathrel{\hbox{\rlap{\hbox{\lower4pt\hbox{$\sim$}}}{\raise2pt\hbox{$<$}}
}}}
\def\ga{\mathrel{\hbox{\rlap{\hbox{\lower4pt\hbox{$\sim$}}}{\raise2pt\hbox{$>$}}
}}}

\newcommand{\ro}{{\sl ROSAT}}
\newcommand{\eu}{{\sl EUVE}}
\newcommand{\iu}{{\sl IUE}}
\newcommand{\hs}{{\sl HST}}

\newcommand{\hip}{{\sl Hipparcos}}
\newcommand{\fuse}{{\sl FUSE}}
\newcommand{\msun}{\mbox{$M_{\odot}\:$}}
\newcommand{\rsun}{\mbox{$R_\odot\:$}}

\newcommand{\nhi}{\mbox{$\rm N_{H\,{\sc i}}\:$}}
\newcommand{\nhei}{\mbox{$\rm N_{He\,{\sc i}}\:$}}
\newcommand{\nheii}{\mbox{$\rm N_{He\,{\sc ii}}\:$}}

\newcommand{\lif}{\mbox{$\rm {LiF}$}}

\newcommand{\tlus}{\mbox{$\rm {\sc tlusty}\:$}}
\newcommand{\iraf}{\mbox{$\rm {\sc iraf}\:$}}

\newcommand{\xsp}{\mbox{$\rm {\sc xspec}\:$}}
\newcommand{\teff}{\mbox{$T_{\rm {eff}}\:$}}
\newcommand{\bcrt}{\mbox{$\beta$\,Crt}}
\newcommand{\bcrtb}{\mbox{$\beta$\,Crt\,B}}
\newcommand{\bcrta}{\mbox{$\beta$\,Crt\,A}}

\begin{document}

\label{firstpage}

\maketitle

\begin{abstract}

We present {\fuse} 
H Lyman series spectroscopy of the hot white dwarf companion to the 
4th magnitude A1\,III star {\bcrt}, which  
shows that is has an unusually 
low mass, $M_{\rm WD}=0.43{\msun}$, and has almost certainly evolved 
through binary interaction. This system could be a long-sought remnant of 
Algol-type evolution, although 
radial velocity measurements appear to show that the pair are not close.  
Instead, micro-variations in {\bcrt}'s proper motion as measured by 
{\hip} suggests that the period could be as high as 
$\sim10$~years. However, a low mass 
white dwarf in a system with a period $\ga3$~years is difficult to 
explain by conventional models for binary evolution.  
We speculate on alternative models for the evolution of this system 
which involve an eccentric binary or 
multiple components.

\end{abstract}

 
\section{Introduction}

Sirius-like binaries contain a main sequence or evolved 
star (spectral type K or earlier) 
and a non-interacting white dwarf secondary. Until recently only a 
handful of such systems were known, such as the prototype Sirius, 
where the white dwarf 
Sirius~B (1.03\msun, Holberg et~al. 1998) is in a wide visual orbit about the 
A0\,V star Sirius~A (P$\sim50$~years). Since then, 
through the {\ro} Wide Field Camera 
(WFC) and the Extreme Ultraviolet Explorer ({\eu}) all sky surveys  of the 
early 1990's, more than 20 new Sirius-like systems have been discovered, 
with primaries as massive as spectral type B (Barstow et~al. 1994, 
Burleigh, Barstow \& Fleming 1997, Burleigh \& Barstow 1998, 
Vennes, Christian \& Thorstensen 1998, hereafter VCT98). 
None of these new systems  are 
spatially resolved from the ground, although 
Barstow et~al. (2001) have recently resolved a number in the far-UV using 
the WFPC2 camera on {\hs}.

Sirius-like binaries are of astrophysical interest because they can be used 
to investigate the relationship between the mass of a main sequence star 
and its white dwarf progeny -- the initial-final mass relation 
(Weidemann 1987). Assuming the pair have never interacted, the white dwarf 
must have evolved from an object more massive than the current primary. 
In addition, 
if the two components can be resolved and an astrometric mass determined 
for the degenerate star, these systems can potentially be used to 
investigate the theoretical white dwarf mass-radius relation, for which 
few data points currently exist (Vauclair et~al. 1997, Provencal et~al. 
1998). 

\begin{figure*}
\vspace{10cm}
\includegraphics{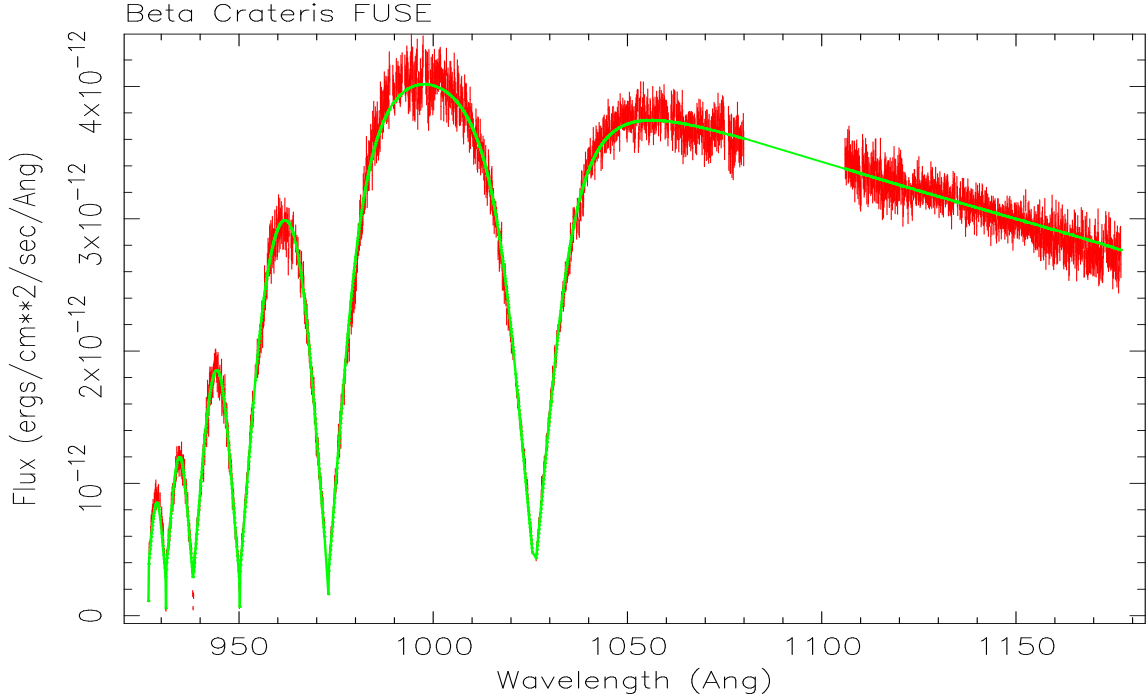}
\caption{FUSE spectrum of the white dwarf companion to Beta Crateris, 
together with a model for $T_{\rm eff}=35,885$K and log $g =7.198$.}
\end{figure*}

The first new Sirius-like system discovered by {\ro} was {\bcrt} 
(HD\,97277 $=$ HR\,4343, A1\,III$+$WD, V$=4.5$, Fleming et~al. 1991). 
Unfortunately, published estimates  
of the effective temperature ({\teff}), surface gravity (log $g$) and 
mass ($M_{\rm WD}$) of the white dwarf in this system  
are poorly constrained. 
Since it is unobservable at optical wavelengths, due to the 
overwhelming brightness of the A1\,III companion,  
we cannot establish these parameters by the usual method of 
fitting a grid of model atmospheres to the optical H Balmer absorption
line series. An unambiguous determination of {\teff} and log $g$
(and, hence, $M_{\rm WD}$) requires three or more spectral lines, 
but existing {\iu} far-UV spectra of the white dwarf 
only cover the H Lyman~$\alpha$ absorption line at 1216{\AA}. 
From a careful analysis of this line's 
profile, Barstow et~al. (1994) and 
VCT98 could only constrain the white dwarf's gravity log $g$ to be 
$\le 8.3$. Assuming both the white dwarf and the A1 giant lie at the
{\hip} distance estimate of $\sim 82$~pc, VCT98 could further constrain
the gravity log $g$ to be $\le 7.6$. This would imply the 
white dwarf has an unusually low mass, 
$M_{\rm WD} \le 0.44 M_\odot$, too low
for it to have evolved in isolation (e.g.,~Marsh, Dhillon \& Duck 1995). 
The progenitor of such a low mass 
degenerate must at some stage 
have interacted with another star, during which it lost mass and thus its 
evolution was accelerated. For example, the progenitor could lose its outer 
envelope without having reached the asymptotic giant branch or ever having 
ignited helium in its core (which requires a core mass of $0.49M_\odot$). 
Nelemans \& Tauris (1998) even speculate that such low mass 
degenerates might form when a solar-like star ascends the giant branch and 
captures a nearby giant planet (e.g.~HD\,89707). The subsequent spiral-in 
phase expels the giant's envelope leaving a low mass helium-core 
white dwarf remnant.

Before we speculate on the possible origins of a low mass white dwarf 
in the {\bcrt} system, we need to be certain that the degenerate star and the 
A1 giant are actually physically related and  
do not merely form a chance alignment. If the white dwarf's gravity was  
in reality nearer log $g=8$, then it would lie in front of the A
giant and have a mass nearer the mean for such objects, $\sim0.6{\msun}$. 
In that case, there would be no need to invoke binary evolution to 
explain its existence. 

The advent of far-UV spectroscopy from 900{\AA} to 1200{\AA} 
with the {\em Far Ultraviolet Spectroscopic 
Explorer} ({\fuse}) satellite, 
covering the entire H Lyman series, now provides us with a first opportunity 
to determine precisely the fundamental parameters of the white dwarf, 
\bcrtb. 
In this paper we show that it   
does indeed have an unusually low mass, and has almost certainly 
evolved through binary interaction.



\begin{figure}
\vspace*{7cm}
\includegraphics{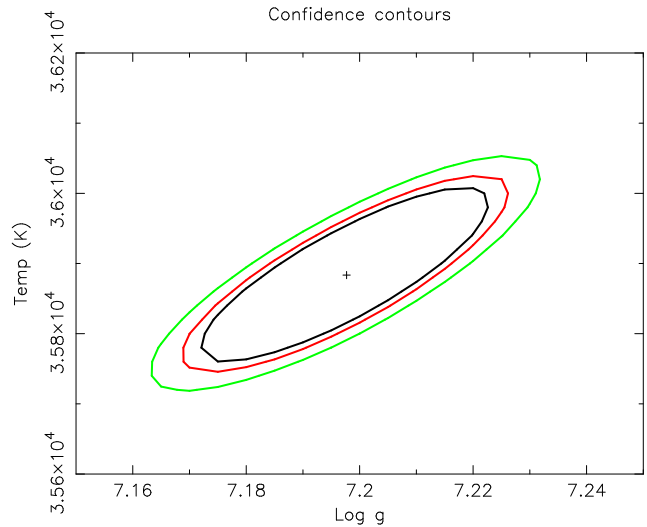}
\caption{Confidence contours for the spectral fit to the 
FUSE data, at 66\%, 90\% and 99\% in the (log $g$, $T_{\rm eff}$) plane.}
\end{figure}

\section{Observations and data reduction}

\subsection{FUSE}

{\fuse} was launched from 
Cape Canaveral on 1999 June 24, and is designed to make high resolution 
spectroscopic observations in the far-UV regime (905{\AA}$-$1187{\AA}), 
in particular covering the H Lyman series from Ly$_{\beta}$ at 1026{\AA} 
to the Lyman limit at 912{\AA}.  
The instrument consists of four 
co-aligned prime focus telescopes and Rowland spectrographs with 
microchannel plate detectors. Two of the telescope channels use 
{\mbox{$\rm {Al}$}}:{\mbox{$\rm {LiF}$}} coatings for optimum reflectivity 
between 1000{\AA}$-$1087{\AA}, and the other two channels use 
{\mbox{$\rm {SiC}$}} coatings for optimized throughput between 
905{\AA}$-$1105{\AA}. Each channel is further subdivided into two 
non-continuous segments (they are split by a few angstr\"oms). 
For a detailed overview of the {\fuse} mission and instruments, 
see Moos et~al. (2000). 

{\bcrt} was observed by {\fuse} for a total of 8037~s on 2000 May 29. 
Two separate exposures were made, each of 4018.55~s. 
At that stage of the mission, {\fuse} 
was experiencing unexpectedly large changes in 
the alignment of the optics during each orbit due to thermal motion. 
In particular, small rotations 
in the mirror made it difficult to keep the four 
independent optical paths within the instrument co-aligned, 
sometimes resulting in misalignments large enough for targets to drift out of 
the aperture in one or more channels (for a full discussion of the on-orbit 
performance of {\fuse}, see Sahnow et~al. 2000). 
The {\bcrt} observations were therefore made 
using the low-resolution aperture (LWRS:
$30\arcsec\times30\arcsec$) in order to minimise these problems,
but none-the-less the flux calibration and wavelength calibration of
different segments of the spectrum were not originally altogether 
satisfactory. Re-extraction of the data with version 1.8.7 of the
CALFUSE pipeline, released to the community in late-2000, resulted in
a marked improvement in the flux calibration for each segment,
although small differences were still noticeable. Consequently, we
have scaled the flux level of each extracted segment to that of the
{\lif}~1a segment, since the {\lif}~1 channel is used for guiding and
is therefore controlled. Even after re-extraction and re-calibration with 
the most recently available wavelength calibration files, differences
in the wavelength calibration of different 
segments of the spectrum were clearly noticable. 
Again, each extracted segment was therefore shifted with respect to
the calibration for the {\lif}~1 channel. Even so, the absolute 
wavelength
calibration for {\lif}~1 is not necessarily reliable, and thus any 
line velocities should be treated with caution.

The two exposures for each segment were co-added, and then each
segment was merged to form one continuous spectrum. Regions of low
signal/noise and/or poorly calibrated data were rejected at this
stage. 
The merged spectra were then 
re-binned to 0.1{\AA} for the purpose of fitting the 
H Lyman lines to determine the star's temperature and gravity, although the 
measured resolution is somewhat better, R$\approx$18000$-$24000 depending 
on the wavelength region.

\subsection{EUVE}

{\eu} observed {\bcrt} for $\sim110,000$~s in 1998 February. 
Continuum flux from 
the white dwarf was detected in both the short wavelength (SW, 70$-$190{\AA}) 
and  medium wavlength (MW, 140$-$350{\AA}) spectrometers. 
Note that the A1\,III primary contributes no flux at these 
wavelengths. We have extracted 
these data from the {\eu} GO-processed event data using standard {\iraf} 
proceedures. Our reduction techinques are described in more detail in 
earlier work (e.g.~Barstow et~al. 1997).

\section{Data analysis}


\subsection{Temperature, gravity and mass from FUSE}

We have matched the {\fuse} spectrum with a grid of non-LTE model
atmospheres, assuming a pure-H composition, 
calculated using the {\tlus} code (e.g.,~Lanz \& Hubeny 1995). 
The grid spans 20,000K to 100,000K in 
{\teff} and 6.5 to 9.75 in log $g$. We fitted five H 
Lyman absorption lines shortwards of and including Ly$_\beta$, 
and the continuum redward 
from Ly$_\beta$ to $\approx1180${\AA}. Geocoronal emission lines in 
the cores of Ly$_\beta$ and Ly$_\gamma$ were ignored in the fit, 
as were regions around narrow ISM absorption lines such as OI at 
1039.2{\AA} and 988.7{\AA}. A model was included in the fit to account 
for the narrow HI ISM absorption lines seen in the core of each H
photospheric line. Although this is 
necessary in order to correctly model the
broad photospheric absorption lines, the HI ISM absorption lines are  
saturated and cannot be used to determine the column density. 
Instead, we have estimated the HI column density to {\bcrt} from 
the {\eu} spectrum, see below. 

A best fit model was found by fitting
our combination of photospheric and ISM models within the program {\xsp}, 
in a similar manner to fitting the H Balmer series
in optical spectra (e.g.,~Marsh et~al. 1997). We find the best fit 
parameters to be {\teff}$=35,885$~K 
(formal 1$\sigma$ errors 35,810~K$-$35,940~K) and log $g$ $=$7.198 
(7.190$-$7.209). The model fit is shown in Figure 1. 
Confidence contours for this fit are shown in Figure 2. 
We note that slight variations on this fitting 
(e.g.~adding or ignoring a few extra data points at the short wavelength end)
can produce slightly different results
with very similar reduced chi-squared values ($\chi_r^2$), 
but that the best fit parameters always lie within the ranges 
35,750K$<${\teff}$<$35,950K and 7.17$<$log $g$$<$7.21. Inclusion of higher 
order H lines results in much poorer fits ($\chi_r^2>2.0$), although the 
surface gravity is slightly reduced as a consequence. This is not
altogether surprising since the higher 
order lines are more contaminated by ISM absorption than the stronger, 
lower order lines. On the other hand, inclusion of the 
continuum longwards of Ly$_\beta$ is important, 
since this includes some of the redward wing of this line and the 
measured gravity is found to be higher by $\sim$0.05~dex 
without these data. 

Extending the best fit model into the optical region, we estimate the
white dwarf's magnitude at V$=13.4\pm0.2$. Of course, this is
dependent on the error on the {\fuse} flux calibration, which we have
assumed to be $\sim10-15\%$ 
(note that the likely error on the flux calibration
is much greater than the formal errors on the best fit {\teff} and log $g$). 
If V$=$13.2, we estimate the absolute
magnitude M$_v$$=8.19$ and d$=$101~pc, compared to the {\hip} distance
estimate to the primary of 77$-$87~pc. We do not regard this difference as 
significant given the uncertainties in the {\fuse} calibration. 

\begin{table}
\begin{center}
\caption{Physical parameters of the white dwarf companion to {\bcrt}, 
from {\fuse} and {\eu} spectroscopy}
\begin{tabular}{cc}
{\teff} & 35,885K (90\% error: 35,810K$-$35,940K) \\
log $g$ & 7.198 (90\% error: 7.190$-$7.209) \\
Mass  & 0.43{\msun}  \\
Radius & 0.027{\rsun}  \\
V & 13.4$\pm$0.2 \\
M$_v$ & 8.19  \\
{\nhi} & $9.25\pm0.25\times10^{18}$~atoms~cm$^{-2}$ \\
{\nhei} & $1.0\times10^{18}$~atoms~cm$^{-2}$ \\
{\nheii} & $0.4\times10^{18}$~atoms~cm$^{-2}$ \\
\end{tabular}
\end{center}
\end{table}


\subsection{Heavy element opacity and ISM columns from EUVE}

We attempted to match the {\eu} spectrum with the same grid of non-LTE 
model atmopheres as used for the {\fuse} data, again assuming a 
pure-H composition, together with an ISM model, in order to constrain the 
interstellar column densities of {\nhi}, {\nhei} and {\nheii}. 
The white dwarf temperature and gravity were fixed within the 90\% limits 
determined from fitting the {\fuse} data, and the ${\nhi}/{\nhei}$ and 
${\nhei}/{\nheii}$ ISM ratios were fixed according to the mean ionisation 
fractions in the local ISM determined by Barstow et~al. (1997).    
Unfortunately, a good fit could not be satisfactorily obtained. The pure-H 
model predicts a factor $\sim2$ more flux below $\sim120${\AA} than is seen 
in the {\eu} data. Therefore, an additional source of opacity is needed. 

Incorporating a homogeneous mixture of H$+$He into the model provided a 
reasonable match to the data for log~${\rm (H/He)}=4.2$ ($\chi_r^2=2.8$). 
A better match ($\chi_r^2<2.0$) was 
obtained using a stratified LTE H$+$He model atmosphere (Koester 1991),  
which would suggest the 
white dwarf's atmosphere consists of a thin-H layer overlying a much thicker 
He envelope. However, it is more likely that the additional opacity is 
provided by a mixture of elements heavier than He, supported within  
the white dwarf against gravitional settling by radiation pressure. A 
similarity can be drawn here with the white dwarf RE~J0720$+$318. Burleigh, 
Barstow \& Dobbie (1997) 
demonstrated that this object's {\eu} spectrum could be 
matched with a stratified H$+$He model atmosphere, suggesting it has a 
thin surface H layer, but Dobbie et~al. (1999) later demonstrated that 
a sophisticated model incorporating a mixture of radiatively levitated 
heavy elements provided a more realistic and convincing description of 
RE~J0720$+$318's atmospheric structure and composition.

A satisfactory match ($\chi_r^2<2.0$) to {\bcrtb}'s 
{\eu} spectrum was obtained with a fully line-blanketed non-LTE 
model incorporating a homogeneous mixture of 
those heavy elements expected to provide most of the 
opacity in this temperature regime - C, N, O and Si (Figure 3). 
The abundances for these 
elements were taken from the predictions of Chayer et~al. (1995), 
which were calculated in the context of an equilibrium radiative 
levitation theory: 
C$/$H$=4.98\times10^{-6}$, N$/$H$=2.47\times10^{-6}$, 
O$/$H$=8.99\times10^{-7}$, Si$/$H$=3.37\times10^{-6}$. Again, the 
white dwarf temperature and gravity were fixed within the 90\% limits 
determined from fitting the {\fuse} data, and the ${\nhi}/{\nhei}$ and 
${\nhei}/{\nheii}$ ISM ratios were fixed according to the mean ionisation 
fractions in the local ISM determined by Barstow et~al. (1997). 



\begin{figure}
\vspace{7cm}
\includegraphics{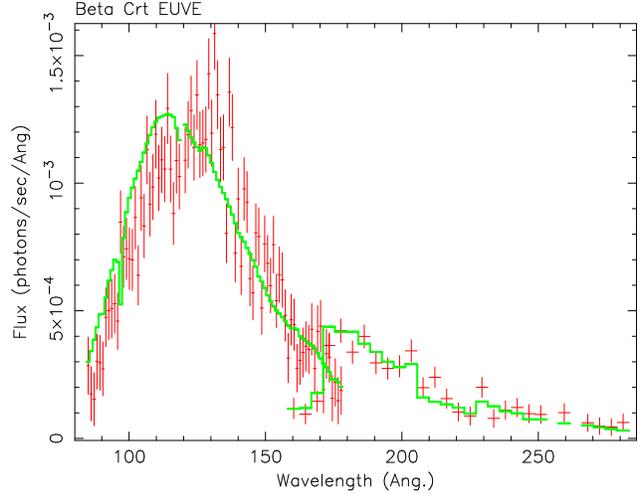}
\caption{EUVE spectrum of the white dwarf companion to Beta Crateris, 
together with a model fit incorporating radiatively levitated 
quantities of C, N, O and Si. See text for details. }
\end{figure}


\subsection{A search for variability in the EUVE light curve}

The hot white dwarf 
GD394 is variable in the EUV with a period of 1.150$\pm$0.003~days 
(Dupuis et~al. 2000). The white dwarf companion to 
V471~Tauri is also variable in the EUV with a period 
of 555~sec. (Jensen et~al. 1996, Dupuis et~al. 1997). In both cases the
variability is probably due to an accretion `spot' 
of heavy elements such as Si on the stellar 
surface. The `spot' appears dark at 
EUV wavelengths due to absorption of emergent flux. 
We have examined {\bcrtb}'s {\eu} Deep Survey Lexan/B ($67-178${\AA}) 
detector light curve for variability with periods ranging from 500s
to ~3 days,  
but find no significant evidence for such behaviour. 


\subsection{A search for heavy element lines in the FUSE spectrum}

Given the detection of a source of opacity in the {\eu} spectrum of 
{\bcrtb}, we have searched the {\fuse} data for absorption lines due to 
elements heavier than He. No such features could be identified, other than 
low ionisation ISM lines e.g.~OI, CII, CIII, NI, NII, NIII and ArI. 

We can, though,  place limits on the photospheric abundances of C, N, O and Si 
in the {\fuse} wavelength range (these are the  
elements in our best-fitting model to the {\eu} data). Spectra were 
computed from the previous best-fit pure-H model to the {\fuse} data 
using the program {\rm SYNSPEC} and incorporating C, N, O and Si at the 
abundances predicted by Chayer et~al. (1995, Table 3), 
and compared to the data. 
Calculations were also carried out for 0.1, 0.01, 0.001, 0.0005 and 0.0002 
times these abundances, maintaining the relative fractions of each element 
constant and identical to the predicted ratios. No O lines are visible  
within the {\fuse} range at the Chayer et~al. predicted abundance. We 
estimate the limiting abundances for C, N and Si in this range at
a few $\times10^{-9}/$H. We note that the presence of C, N, O and Si at 
these abundances would have little effect on our determination
of the effective 
temperature and surface gravity (Barstow, Hubeny \& Holberg 1998). 

\section{Discussion}

\subsection{The white dwarf mass}

{\fuse} spectroscopy of the H Lyman series of the hot white dwarf companion to 
{\bcrt} constrains its effective temperature to ${\teff}\approx35,900$K and 
surface gravity log $g\approx7.20$. Combining these measurements with 
the evolutionary models of Bloecker \& Driebe (Bloecker 1995, Dreibe et~al. 
1998) for He- and CO-core white dwarfs, 
we find $M_{WD}=0.43~M_\odot$, confirming it does indeed 
possess an unusually low mass, although it lies on the boundary between 
their He- and CO-core models with a cooling age $\sim10^6$~years.
This result 
suggests that it must have evolved through binary interaction, 
during which the white dwarf progenitor lost mass and its evolution was 
accelerated. 

\subsection{Heavy element abundances}

Model fits to {\bcrtb}'s {\eu} spectrum require an additional source of 
atmospheric opacity, other than simply assuming a pure-H composition. Although 
a satisfactory fit could be achieved with a stratified model which assumes a 
thin surface H layer overlying a thicker He envelope, a good fit could also 
be achieved with a sophisticated non-LTE model incorporating trace abundances 
of radiatively levitated C, N, O and Si. 

We note that 
it was previously believed that the photospheres of 
white dwarfs below ${\teff}\approx40,000$K  
were essentially pure-H in composition. In fact, at the low surface gravity 
of {\bcrtb} it is possible for these relatively high 
heavy element abundances to be 
radiatively levitated within the atmosphere 
at a temperature as low as ${\teff}\approx36,000$K, 
supplying the opacity seen in the {\eu} spectrum (see Table 2 of 
Barstow et~al. 1997). 

Comparisons can be drawn 
with the apparantly 
isolated white dwarf GD394 and the white dwarf secondary in the 
V471~Tauri binary system. Both stars are in a similar 
temperature regime to {\bcrtb} and 
show strong evidence for heavy elements in their atmospheres, but in contrast 
to {\bcrtb} they have much higher surface gravities (e.g.,~for V471~Tauri 
log~$g\approx8.3$, Barstow et~al. 1997b) and 
the heavy elements are probably being accreted rather than simply 
levitated and maintained by radiation pressure. 
For example, GD394 possesses an unusually high Si abundance for a 
${\teff}\approx39,000$K white dwarf 
(Dupuis et~al. 2000). Most likely, for this object, 
the material is being 
episodically accreted, 
either from the ISM or from an unseen low mass companion. The 
white dwarf secondary in the V471~Tauri system has ${\teff}\approx32,000$K 
(Barstow et~al. 1997b), and the indications are that this object is also 
actively accreting Si, in this case from its close, active K2\,V companion 
(Sion et~al. 1998).  

The lack of any variability in {\bcrtb}'s EUV light curve argues 
against accretion as the source of the heavy elements in {\bcrtb}'s 
atmosphere, since these atmospheric contaminents do not appear to be 
concentrated at any particular surface ``spot'', unlike GD394 
(Dupuis et~al. 1997), V471~Tau (Jensen et~al. 1996) and 
RE~J0720$-$318 (Dobbie et~al. 1999). {\bcrtb} is, 
therefore, one of the coolest DA white dwarfs to possess radiatively 
levitated heavy elements in its photosphere. 

The non-detection of C, N and Si lines in the far-UV in the {\fuse} spectrum 
at even a fraction of the 
Chayer et~al. predicted abundances suggests that the white dwarf atmosphere is 
either stratified, with these heavy elements lying below the far-UV line 
forming region or, alternatively, that  
there is an different source of heavy element opacity, e.g.~Fe. 
Unfortunately, since the A giant dominates the white dwarf 
flux above $\sim$1300{\AA} it will always be impossible to search for 
evidence of heavy elements at longer UV wavelengths
with e.g. HST/STIS-echelle, and our best hope of finding any heavy element 
absorption lines remains high resolution, high signal-to-noise 
{\fuse} observations.

\subsection{The association of the white dwarf and {\bcrt}}

Before considering any evolutionary history for the {\bcrt} system we must, 
of course, be convinced that the A1 giant and the white dwarf 
do indeed form a physical pair. Noting that Barstow et~al. (2001) failed to 
resolve the two stars with HST/WFPC2, then their maximum separation is just 
0.08$\arcsec$. Fleming et~al. (1996) gives the space density of soft x-ray 
bright white dwarfs as $\sim1\times10^{-5}$ per cubic parsec. Assuming a 
distance of $\sim100$pc for the white dwarf (as estimated from its  
magnitude in section 3.1), we calculate 
that the chance of a hot white dwarf randomly falling within any area of the 
sky with diameter 0.08$\arcsec$ is $\sim1\times10^{-11}$. 
We therefore regard the chance alignment of {\bcrt} and the white dwarf  
as highly unlikely. We emphasise that although our 
model fit to the white dwarf's {\fuse} spectrum results in a distance estimate 
slightly larger than the {\hip} distance of 77$-$87~pc to 
the A1 giant, we do not regard the difference 
as significant due to the uncertainties in the {\fuse} flux calibration.

\subsection{The binary period}

Sporadic measurements of {\bcrt}'s radial velocity early in the 20th 
Century by Campbell \& Moore (1928) seemed to indicate that it is variable, 
with a full amplitude $\sim20$~km~s$^{-1}$. This led Fleming et~al. (1991) 
to speculate on a possible orbital period $<20.1$~days. 
Smalley et~al. (1997) rejected this period, but a   
re-analysis of Campbell \& Moore's data still showed that the binary 
period could be $\la160$~days. However, Smalley et~al.'s own radial velocity
measurements, and those of Duemmler et~al. (1997) and VCT98 contradict 
Campbell \& Moore's data. These studies indicate that the orbital period of 
{\bcrt} is more likely to be measured in years, if indeed the velocity is 
varying at all. For example, 
Smalley et~al. (1997) observed practically no orbital motion between 
1994 March and 1995 January, giving a full amplitude 
$K=1.9\pm1.3$~km~s$^{-1}$. In contrast,  
Duemmler et~al. (1997) found that the radial velocity they measured 
in 1997 May is different by more than 8~km~s$^{-1}$ from those measured by 
Smalley et~al. 2-3~years earlier. Duemmler et~al. could not 
derive a period from these data, but both groups are in  
agreement that the velocity is stable over periods of days to a year.

Additional velocity measurements by Grenier et~al. (1999) obtained in the 
early 1990's also appear to show variation. For example, 
in 1990 January they measured the velocity at $+7.4\pm0.5$~km~s$^{-1}$, but a 
year later (1991 March) it had apparantly dropped to 
$+2.5\pm0.6$~km~s$^{-1}$. By 1993 March the radial velocity was back at 
$+6.6\pm0.4$~km~s$^{-1}$. 

If all of these measurements and their errors 
are believed, and Duemmler et~al. in particular carefully consider the errors 
in their data, then it is plausible that the 
radial velocity is slowly 
varying between observations taken a few years apart.

On the other hand, 
VCT98 monitored {\bcrt} for over 15 months from 
1995 December to 1997 March, covering the period between the Smalley et~al. 
and Duemmler et~al. observations, and saw no obvious velocity variations 
during this time. They concluded that there are only 
slight deviations from a systemic velocity of 
$\gamma=+8.6\pm1.6$~km~s$^{-1}$, limiting the  velocity 
amplitude $K<1.6$~km~s$^{-1}$. 


Clearly, then, the {\bcrt} system deserves regular 
monitoring in order to confirm firstly whether there are any long-term radial 
velocity variations due to orbital motion at all 
and, if there are, then to categorically determine
 the binary period and the velocity amplitude.   


Perhaps a better indication of the binary period arises from micro-variability 
in the proper motion of {\bcrt} as measured by {\hip}. These data indicate 
that any  
binary period is probably $\sim$10~years. This would be consistent with the 
failure of Barstow et~al. (2001) 
to resolve the system in the far-UV with {\hs} and WFPC2, limiting  
the separation 
to $<0.08\arcsec$ and placing an upper limit on the binary period of 
$\sim9.5$~years, assuming a favourable geometry. From this diverse   
evidence we conclude that 
the {\bcrt} system is not a close binary, unless the system is being viewed at 
a very unfavourable geometry (i.e.~face-on). Further, it is plausible 
that the period is of order years, and that it may be as high as 
$\sim10$~years.

\subsection{A post-Algol system?}

Low mass ($M \la 0.5 M_\odot$), helium-core white dwarfs cannot be made by 
standard single star evolution since the main sequence lifetime of
their progenitor stars is longer than the Hubble time. 
The usual scenario for making such white dwarfs involves a 
phase of mass transfer between stars in a binary system.
In this picture the expansion of the white dwarf progenitor, 
as it evolves up the giant branch, causes it to fill its Roche lobe 
while the core mass remains low. Mass transfer to the 
secondary star (in this case $\beta$~Crt~A) then exposes the low mass, 
helium core.  

The {\bcrt} system could, therefore, be a long-sought 
remnant of Algol-type evolution. In an Algol-type binary, a cool 
F-K\,III-IV secondary star fills its Roche lobe and transfers mass to 
a  hot B-A\,V primary. 
These systems were a puzzle to astronomers because the 
less massive secondary is more evolved than its companion. This paradox was 
resolved when it was realised that the current secondary was originally 
the more massive star, and that the current configuration of the binary was 
the result of extensive mass transfer, resulting in a reversal of the mass 
ratio of the system. A remnant of such an Algol-type binary 
would then consist of an early-type 
intermediate-mass star with a low mass white dwarf 
companion. Although Algols are the most numerous of known eclipsing binaries, 
their post-mass transfer counterparts have eluded discovery because the 
white dwarfs are overwhelmed at optical wavelengths by their bright 
companions. {\bcrt} could be the first confirmed post-Algol system. 

If {\bcrt} is indeed a remnant Algol-type binary, we would expect the orbital 
period to be 
$\la$~tens of days. Of course, as discussed above, the period is currently 
unknown. It is possible that the system appears face-on to us, so that it 
shows little or no radial velocity variations, even if the period was short. 
If, on the other hand, the period is $\ga$years, then this would start to 
cause problems for standard models of binary evolution.

\subsection{Alternative evolutionary scenarios}

As discussed in Section~4.4 above, there is evidence to suggest 
that the orbital 
period is in fact relatively long, perhaps as high as $\sim10$~years. 
Unfortunately, it is difficult to explain the existence of such a system 
with standard binary evolution models. For example, 
the Roche lobe radius of the white dwarf 
progenitor in, say, a 7~year orbit is $\ga260R_{\odot}$. 
The only stars with radii this large have masses of 
$\ga 10M_\odot$, and are on the giant branch
with helium core masses much larger than $0.4M_{\odot}$. 
In fact, the largest radius at which a star
of any mass has a helium core of $0.4M_{\odot}$ is about $100
R_{\odot}$, for a $0.8M_{\odot}$ star almost at the tip of the giant
branch (Webbink 1985). 
More massive stars create helium cores of $0.4M_{\odot}$ much
earlier in their evolution, and hence at smaller radii. 
Stars with masses greater than about $4M_{\odot}$ reach core masses
of $0.4M_{\odot}$ on the main sequence. Therefore, it seems clear
that the white dwarf progenitor must have been less
massive than $4M_{\odot}$. However, the  large Roche radius of this 
progenitor star leaves the 
standard scenario facing a number of severe difficulties in
explaining the existence of a low mass white dwarf in a system 
with an orbital period as high as $\sim10$~years. 
Nuclear-driven, conservative, stable mass transfer from
the white dwarf progenitor to the secondary while the progenitor was on 
the giant branch could increase the orbital period of the system 
(Webbink, Rappaport \& Savonije 1983; Ritter 1999), but 
it is not possible to increase the 
orbital period beyond $\sim3$~years for a helium-core white dwarf of 
$0.4M_\odot$. Therefore, 
the possible $\sim10$ year orbital period of {\bcrt}, 
or indeed any orbital period significantly longer than $\sim3$~years,  
remains a puzzle in the context of conventional models of binary
evolution. 

We now consider a number of other evolutionary  possibilities. \\
\newline
\noindent
(a) {\em Unstable mass transfer.}
If the
initial binary system had an extreme mass ratio, or if the white dwarf
progenitor had a deep convective envelope when mass transfer was
initiated, the mass transfer could have become unstable and occur on the
thermal or dynamical timescale of the mass donor. Under these 
circumstances, the
analytic and semi-analytic models break down, and it is difficult to
determine the outcome of such a situation. It is possible
that the white dwarf progenitor could be driven far out of
equilibrium, allowing it to 
continue to fill its Roche lobe as the mass ratio reverses and the
orbit expands to the observed period. Calculations of thermal
timescale mass transfer (Schenker 2001),
in which the donor star is out of thermal equilibrium, suggest that
the system behaves differently for a short time. However, it returns
to a stable mass transfer configuration on the giant branch reasonably
quickly. On the other hand, dynamical 
instability of the donor star is usually assumed to lead to common 
envelope (CE) evolution. While CE evolution is a viable mechanism for
the production of low mass white dwarfs it is also associated with a decrease
in the orbital period, as the orbital angular momentum is reduced in
driving off the envelope. Therefore, we conclude that it is  
unlikely that a period of unstable mass transfer or
non-equilibrium evolution of the white dwarf progenitor alone could 
produce the low mass white dwarf in  {\bcrt}.\\
\newline
\noindent
(b) {\em Period evolution after mass transfer ceases.}
A long period system could be explained by 
allowing the orbital period to increase after the 
white dwarf has been formed. In general, a binary system can 
increase its period if mass is lost from
the system with a relatively low specific angular momentum. 
The obvious physical example of such mass loss would be the
presence of an intense stellar wind from one or both of the
stars in {\bcrt}. However the mass loss would have to be extreme: 
we could envisage a system in which {\bcrta} was initially much
more massive. Some B stars are known to have significant winds, and
the mass lost through such a wind could be lost from the binary
system. The observed mass loss rates for B stars ($10^{-5}$ -- $10^{-7}
M_{\odot}$ yr$^{-1}$, Lamers 1981) are such that it would
take $\sim10^7$ years to turn a $25M_{\odot}$ B star into a $4
M_{\odot}$ A star. Even if this extreme mass reduction was allowed
within the main sequence life of such a star the cooling age of 
the white dwarf is only $\sim10^6$ years, an order of 
magnitude shorter than the required mass loss timescale. 
So it would seem that the large magnitude of the period increase 
and the short timescale on which it needs to act combine to make the 
period increase solution very unlikely. \\
\newline
\noindent
(c) {\em A multiple system.}
It is possible that {\bcrt} is in fact a triple system, with 
{\bcrta} orbiting a short period binary containing the white dwarf 
{\bcrtb} and another star. In this case the evolution of the shorter
period binary system resulted in the creation of the 
low mass white dwarf. The
progenitor of the white dwarf must have been more massive than 
{\bcrta}, which is assumed to be unaffected by the evolution of the
smaller binary system. Hence the age of the system is set by
the age of the A1 giant {\bcrta}. 
In addition, the white dwarf has a cooling age
of $\sim 10^6$ years, which means that the progenitor has to be only
slightly more massive than A1\,III.  Since the companion to the
white dwarf is currently unseen, its mass must be less than 
$\sim0.8M_{\odot}$. In order to produce a white dwarf mass of $0.4M_{\odot}$,
a $3.5M_{\odot}$ star must have filled its Roche lobe with a radius of
about $10^{12}$ cm. From these constraints on the masses of the two
binary components and the Roche lobe radius, we calculate that such a
system would have an orbital period of about 6 hours. It would be
interesting to measure the radial velocity of the white dwarf itself
to see if it varies between the two {\it FUSE} exposures, but the lack
of any narrow heavy element lines, the infilling of the H lines with
geocoronal emission and/or interstellar absorption, and the poor
wavelength calibration precludes any such measurement with these data.\\
\newline
\noindent
(d) {\em An eccentric binary system.}
Another possibility involves an eccentric binary system, in
which mass transfer from the primary occurred in bursts at
periastron. If the binary system had an eccentricity of $\sim0.5$ and
a period $\sim10$~years, then the Roche lobe radius of {\bcrtb} at
periastron is consistent with the radius of a giant that has a core mass
of $0.4M_{\odot}$. Hence it is possible 
that the {\bcrt} system was formed from a binary system
consisting of two stars of similar mass ($\sim2M_{\odot}$) in
a wide, eccentric orbit. The more massive of these stars began
evolving up the giant branch. Only when its core mass was greater than
about $0.25M_{\odot}$ did the star fill its Roche lobe and begin
transferring mass (conservatively) to the lower mass star. Since the
system had an eccentric orbit, mass transfer only occurred at
periastron. This changed the eccentricity of the orbit (Matese \& 
Whitmore 1983), but clearly the final system must have had parameters which
meant that the Roche lobe radius was $\sim100 R_{\odot}$ and the  
orbital period was of $\sim10$ years. 

\section{Conclusion}

{\fuse} spectroscopy of the hot white dwarf companion to {\bcrt} shows 
that it has an unusually low mass ($M_{\rm WD}=0.43{\msun}$)  
and has almost certainly evolved through binary 
interaction. Thus, the system could be a long-sought remnant of Algol-type 
evolution. However, radial velocity measurements appear to indicate that the 
pair are not close. If the velocity is essentially static then we could 
of course be viewing the system at an unfavourable geometry (i.e.~face-on). 
Alternatively, the binary 
is relatively wide. If the variations between radial 
velocity measurements made by Grenier et~al. (1999) in the early 1990s, 
Smalley et~al. (1997) in 1994/95, and Duemmler 
et~al. (1997) in 1997 May are believed, then the period could be of order 
years. Failure to resolve the 
pair with HST/WFPC2 places an upper limit on the period of 
$\la9.5$~years, assuming a favourable geometry. Perhaps more importantly, 
micro-variations in 
{\bcrta}'s proper motion as measured by {\hip} suggest a binary period 
$\sim10$~years. But any binary period $\ga3$~years poses a problem for 
standard models of binary interaction. 
The most likely evolutionary history of the $\beta$~Crt system 
which could result in a low mass white dwarf in a $\sim10$ year orbit with an
A1 giant is one which involves either a multiple system or an
eccentric binary orbit. Clearly, though, 
the orbital period urgently needs to be 
determined before we can distinguish between these possible 
evolutionary scenarios. 


\section{Acknowledgements} 

This work  
was supported by PPARC, UK\null. MRB would like to thank B-G Andersson and 
all the members of the {\fuse} team at Johns Hopkins University, Baltimore, 
for their help in reducing, understanding and analysing these data. 
The {\rm TLUSTY} and {\rm SYNSPEC} codes 
were developed by Ivan Hubeny. 
This research has made use of the {\it SIMBAD} database operated by CDS, 
Strasbourg, France.

\end{document}